\documentclass[12pt]{article}

\usepackage{graphicx}
\begin{document}

\begin{center}
{\bf Dyonic and magnetized black holes based on nonlinear electrodynamics} \\
\vspace{5mm} S. I. Kruglov
\footnote{E-mail: serguei.krouglov@utoronto.ca}
\underline{}
\vspace{3mm}

\textit{Department of Physics, University of Toronto, \\60 St. Georges St.,
Toronto, ON M5S 1A7, Canada\\
Department of Chemical and Physical Sciences, University of Toronto,\\
3359 Mississauga Road North, Mississauga, ON L5L 1C6, Canada} \\
\vspace{5mm}
\end{center}
\begin{abstract}
We propose a new model of nonlinear electrodynamics with two dimensional parameters. The phenomenon of vacuum birefringence, the principles of causality and unitarity were studied. It was shown that there is no a singularity of the electric field in the center of point-like charges. Corrections to the Coulomb law as $r\rightarrow\infty$ were obtained. Dyonic and magnetized black holes are considered. We show that in the self-dual case, when the electric charge equals the magnetic charge, corrections to Coulomb's law and Reissner$-$Nordstr\"{o}m solutions are absent. The metric function and its asymptotic as $r\rightarrow\infty$ for magnetic black holes were evaluated. We calculate the magnetic mass of the black hole which is finite. The thermodynamic properties and thermal stability of black holes were investigated. We calculated the Hawking temperature, the heat capacity and the Helmholtz free energy of black holes and shown that at some parameters there are second-order phase transitions. It was demonstrated that at some range of parameters the black holes are stable.
\end{abstract}

\section{Introduction}

Nowadays, physics of black holes (BHs) attracts much attention. Especially the cases of magnetically and electrically charged BHs are of interest. Dyonic BH (with magnetic and electric charges) solutions were obtained in the string action \cite{Shapere}-\cite{Jatkar} and in the theory of supergravity
\cite{Chamseddine}-\cite{Pang}. Dyonic BH solutions have applications in the theory of superconductivity and thermodynamics. In \cite{Hartnoll} the Hall conductivity in the framework of AdS/CFT correspondence was studied, and the Nernst effect was described \cite{Hartnoll1}. Thermodynamic properties of dyonic BH, its dual field theory, phase transitions and other critical phenomena were investigated in \cite{Dutta}. These and other investigations show the importance of studying dyonic BHs.

In this paper we study dyonic and magnetic BHs based on nonlinear electrodynamics (NED). NED is used to solve the problems of singularities in the center of charges and the infinite self-energy problem. The first model of NED that can solve problems of singularities was proposed by Born and Infeld (BI) \cite{Born}. It was shown that in QED quantum corrections produce NED \cite{Heisenberg}.
Other NED models \cite{Soleng}-\cite{Kruglov1} also can solve the singularity problems. NED in general relativity (GR) was considered in \cite{Pellicer}-\cite{Quiros} and thermodynamics of BHs was investigated \cite{Hendi1}-\cite{Krug6}. The phase transitions in electrically and magnetically charged BHs were investigated in \cite{Yajima}-\cite{log}. It worth noting that the universe inflation can be described by NED coupled with GR \cite{Garcia}-\cite{Kruglov4}.

The effect of vacuum birefringence occurs in QED, taking into account loop corrections, when the external magnetic field is present. This phenomenon is of experimental interest \cite{Rizzo}-\cite{Battesti}.
In BI electrodynamics the effect of birefringence is absent but in BI-type electrodynamics with two parameters the birefringence phenomenon holds \cite{Krug}. Thus, models of NED that produce the phenomenon of vacuum birefringence are of definite interest.

In Sec. 2 we propose a new model of NED with two parameters $\beta$ and $\gamma$. The effect of vacuum birefringence is studied. It is shown that at $\gamma =2\beta$ the birefringence phenomenon is absent.
We obtain the range of electromagnetic fields when the causality and unitarity principles hold. It was demonstrated in Sec. 3 that the dual symmetry is violated. We show that a singularity of the electric field in the center of point-like charges is not present and the maximum of the electric field in the center is $E(0)=1/\sqrt{\beta}$. We obtain the corrections to Coulomb's law that are in the order of ${\cal O}(r^{-6})$. In Sec. 4 we show that the scale invariance is broken due to the presence of dimensional parameters. We obtain the dyonic solution in Sec. 5.
In Sec. 6 we study the magnetic BH, and the mass, the metric function and their asymptotic as $r\rightarrow\infty$  are found. It was demonstrated that the magnetic mass of BHs is finite. The BH thermodynamics and the thermal stability are investigated in Sec. 7. We obtain the Hawking temperature, the heat capacity, the Helmholtz free energy and show the possibility of phase transitions in BHs. Sec. 8 is a conclusion.

We use units with $c=\hbar=1$ and the metric signature $\mbox{diag}(-1,1,1,1)$.

\section{The model of nonlinear electrodynamics}

We propose the NED with the Lagrangian density
\begin{equation}
{\cal L} = -\frac{{\cal F}}{\sqrt{2\beta{\cal F}+1}}+\frac{\gamma}{2}{\cal G}^2,
 \label{1}
\end{equation}
where the parameters $\beta$ and $\gamma$ possess the dimensions of (length)$^4$, ${\cal F}=(1/4)F_{\mu\nu}F^{\mu\nu}=(B^2-E^2)/2$, ${\cal G}=(1/4)F_{\mu\nu}\tilde{F}^{\mu\nu}=\textbf{E}\cdot \textbf{B}$, $F_{\mu\nu}=\partial_\mu A_\nu-\partial_\nu A_\mu$ is the field tensor, and $\tilde{F}^{\mu\nu}=(1/2)\epsilon^{\mu\nu\alpha\beta}F_{\alpha\beta}$
is the dual tensor.
Let us study the effect of vacuum birefringence in the model under consideration.
Theoretically, the phenomenon of the vacuum birefringence holds in QED due to loop quantum corrections \cite{Heisenberg}. But this effect is very weak to be verified now in the experiments \cite{Rizzo}, \cite{Valle}.
The Taylor series of the Lagrangian density (1) at $\beta {\cal F}\ll 1$ is given by
\begin{equation}
{\cal L} = -{\cal F}+\beta {\cal F}^2-\frac{3}{2}\beta^2{\cal F}^3+\frac{\gamma}{2}{\cal G}^2+{\cal O}\left((\beta {\cal F})^4\right).
 \label{2}
\end{equation}
In our model the correspondence principle holds because the Lagrangian density (1) becomes the Maxwell
Lagrangian density ${\cal L}_{M}=- {\cal F}$ for weak electromagnetic fields, $\beta {\cal F}\ll 1$.
 The Lagrangian density studied in \cite{Kruglov9} reads
\begin{equation}
{\cal L} =- \frac{1}{2}\left(\textbf{B}^2-\textbf{E}^2\right)+a\left(\textbf{B}^2-\textbf{E}^2\right)^2
+b\left(\textbf{E}\cdot\textbf{B}\right)^2.
\label{3}
\end{equation}
Comparing Eqs. (2) and (3), up to ${\cal O}\left((\beta{\cal F})^2\right)$, we find that $4a=\beta$,
$b=\gamma/2$. According to \cite{Kruglov9} the indices of refraction $n_\perp$, $n_\|$ for two polarizations, perpendicular and parallel to the external magnetic induction field $\bar{B}$, are  given by
\begin{equation}
n_\perp=1+4a\bar{B}^2=1+\beta\bar{B}^2,~~~~n_\|=1+b\bar{B}^2=1+\frac{\gamma}{2}\bar{B}^2.
\label{4}
\end{equation}
Thus, one has the effect of vacuum birefringence when $n_\perp\neq n_\|$.
In accordance with the Cotton-Mouton (CM) effect \cite{Battesti} the difference in the indices of refraction is
\begin{equation}
\triangle n_{CM}=n_\|-n_\perp=k_{CM}\bar{B}^2.
\label{5}
\end{equation}
Making use of Eqs. (4) and (5) we obtain the CM coefficient $k_{CM}=\gamma/2-\beta$.
The experiments gave the bounds
\[
k_{CM}=(5.1\pm 6.2)\times 10^{-21} \mbox {T}^{-2}~~~~~~~~~~(\mbox {BMV}),
\]
\begin{equation}
k_{CM}=(4\pm 20)\times 10^{-23} \mbox {T}^{-2}~~~~~~~~~~~~(\mbox {PVLAS}).
\label{6}
\end{equation}
 One finds the bound on the parameters $\gamma/2-\beta\leq(4\pm 20)\times 10^{-23} \mbox {T}^{-2}$ from the PVLAS experiment. In the case of $\gamma=2\beta$ the phenomenon of vacuum birefringence disappears.
 The bound on the CM coefficient in QED is $k_{CM}\leq 4.0\times 10^{-24}\mbox {T}^{-2}$ \cite{Rizzo}.

\subsection{The causality and unitarity principles}

The general principles of causality and unitarity have to be satisfied. The causality principle says that the group velocity of excitations over the background should be less than the speed of the light. In this case  tachyons are absent in the theory. The unitarity principle guarantees the absence of ghosts. The principles of causality and unitarity require the inequalities \cite{Shabad2}
\[
 {\cal L}_{\cal F}\leq 0,~~~~{\cal L}_{{\cal F}{\cal F}}\geq 0,~~~~{\cal L}_{{\cal G}{\cal G}}\geq 0,
\] \begin{equation}
{\cal L}_{\cal F}+2{\cal F} {\cal L}_{{\cal F}{\cal F}}\leq 0,~~~~2{\cal F} {\cal L}_{{\cal G}{\cal G}}-{\cal L}_{\cal F}\geq 0,
\label{7}
\end{equation}
where ${\cal L}_{\cal F}\equiv\partial{\cal L}/\partial{\cal F}$, ${\cal L}_{\cal G}\equiv\partial{\cal L}/\partial{\cal G}$.
With the help of Eq. (1) we find
\[
{\cal L}_{\cal F}= -\frac{\beta{\cal F}+1}{(1+2\beta{\cal F})^{3/2}},~~~~ {\cal L}_{{\cal G}{\cal G}}=\gamma,~~~~
2{\cal F}{\cal L}_{{\cal G}{\cal G}}-{\cal L}_{{\cal F}}=2{\cal F}\gamma+\frac{1+\beta{\cal F}}{(1+2\beta{\cal F})^{3/2}},
\]
\begin{equation}
{\cal L}_{\cal F}+2{\cal F} {\cal L}_{{\cal F}{\cal F}}=\frac{\beta{\cal F}-1}{(1+2\beta{\cal F})^{5/2}},~~~~
{\cal L}_{{\cal F}{\cal F}}=\frac{\beta(2+\beta{\cal F})}{(1+2\beta{\cal F})^{5/2}}.
\label{8}
\end{equation}
Making use of Eqs. (7) and (8), the principles of causality and unitarity require that $-1\leq\beta{\cal F}\leq 1$ (at $\gamma=0$). In the case $\textbf{E}=0$ we have the restriction $|\textbf{B}|\leq \sqrt{2/\beta}$.  If $\textbf{B}=0$ one has the restriction $|\textbf{E}|\leq \sqrt{2/\beta}$.

\section{Field equations}

In this section we consider flat space-time. The equations of motion are given by
\begin{equation}
\partial_\mu\left({\cal L}_{\cal F}F^{\mu\nu} +{\cal L}_{\cal G}\tilde{F}^{\mu\nu} \right)=0.
\label{9}
\end{equation}
Making use of Eqs. (1) and (9) we obtain field equations
\begin{equation}
 \partial_\mu\left(\frac{-(\beta{\cal F}+1)F^{\mu\nu}}{(1+2\beta{\cal F})^{3/2}}
+\gamma{\cal G}\tilde{F}^{\mu\nu}\right)=0.
\label{10}
\end{equation}
The electric displacement field is $\textbf{D}=\partial{\cal L}/\partial \textbf{E}$,
\begin{equation}
\textbf{D}=\frac{1+\beta{\cal F}}{(1+2\beta{\cal F})^{3/2}} \textbf{E}+\gamma {\cal G}\textbf{B}.
\label{11}
\end{equation}
The magnetic field is given by $\textbf{H}=-\partial{\cal L}/\partial \textbf{B}$,
\begin{equation}
\textbf{H}= \frac{1+\beta{\cal F}}{(1+2\beta{\cal F})^{3/2}}\textbf{B}-\gamma{\cal G}\textbf{E}.
\label{12}
\end{equation}
We use the decomposition of Eqs. (11) and (12) as follows \cite{Hehl}:
\begin{equation}
D_i=\varepsilon_{ij}E^j+\nu_{ij}B^j,~~~~H_i=(\mu^{-1})_{ij}B^j-\nu_{ji}E^j.
\label{13}
\end{equation}
From Eqs. (11), (12) and (13) one obtains
\[
\varepsilon_{ij}=\delta_{ij}\varepsilon,~~~~(\mu^{-1})_{ij}=\delta_{ij}\mu^{-1},~~~~\nu_{ji}=\delta_{ij}\nu,
\]
\begin{equation}
\varepsilon=\frac{1+\beta{\cal F}}{(1+2\beta{\cal F})^{3/2}},~~~~
\mu^{-1}=\varepsilon=\frac{1+\beta{\cal F}}{(1+2\beta{\cal F})^{3/2}},~~~~\nu=\gamma {\cal G}.
\label{14}
\end{equation}
Using Eqs. (11) and (12), field equations (10) can be represented as the Maxwell equations
\begin{equation}
\nabla\cdot \textbf{D}= 0,~~~~ \frac{\partial\textbf{D}}{\partial
t}-\nabla\times\textbf{H}=0.
\label{15}
\end{equation}
Because $\varepsilon_{ij}$, $(\mu^{-1})_{ij}$, and $\nu_{ji}$ depend on electromagnetic fields, Eqs. (15) are nonlinear Maxwell's equations.
Making use of the Bianchi identity $\partial_\mu \tilde{F}^{\mu\nu}=0$, we find the second pair of Maxwell's equations
\begin{equation}
\nabla\cdot \textbf{B}= 0,~~~~ \frac{\partial\textbf{B}}{\partial
t}+\nabla\times\textbf{E}=0.
\label{16}
\end{equation}
From Eqs. (11) and (12) one obtains the equation
\begin{equation}
\textbf{D}\cdot\textbf{H}=(\varepsilon^2-\nu^2)\textbf{E}\cdot\textbf{B}+2\varepsilon\nu{\cal F}.
\label{17}
\end{equation}
Because $\textbf{D}\cdot\textbf{H}\neq\textbf{E}\cdot\textbf{B}$, the dual symmetry is broken in our model \cite{Gibbons}. In classical electrodynamics and in BI electrodynamics the dual symmetry occurs but in QED and  generalized BI electrodynamics \cite{Krug} the dual symmetry is violated.

\subsection{Point-like charges and their fields}

In the electrostatics, the electric displacement field for the point-like particle ($\textbf{B}=0$), in Gaussian units, obeys the equation
\begin{equation}
\nabla\cdot \textbf{D}=4\pi q_e\delta(\textbf{r}),
\label{18}
\end{equation}
where $q_e$ is the electric charge.
The solution to Eq. (18), using Eq. (11), is given by
\begin{equation}
\frac{E\left(1-\beta E^2/2\right)}{(1-\beta E^2)^{3/2}}=\frac{q_e}{r^2}.
\label{19}
\end{equation}
When $r\rightarrow 0$ the solution to Eq. (19) reads
\begin{equation}
E(0)=\sqrt{\frac{1}{\beta}}.
\label{20}
\end{equation}
There is no singularity of the electric field in the center of point-like charges.
The field (20) is the maximum of the electric field at the origin of charged particles. The similar feature holds
in BI electrodynamics. In classical electrodynamics the electric field possesses the singularity in the center of charges. Let us introduce unitless variables
\begin{equation}
x=\frac{\sqrt{2} r^2}{q_e\sqrt{\beta}},~~~~y=\sqrt{\frac{\beta}{2}}E.
\label{21}
\end{equation}
Equation (19) in the terms of unitless variables (21) is written as
\begin{equation}
\frac{(1-2y^2)^{3/2}}{y(1-y^2)}=x.
\label{22}
\end{equation}
The function $y(x)$ is depicted in Fig. 1.
\begin{figure}[h]
\includegraphics[height=3.0in,width=3.0in]{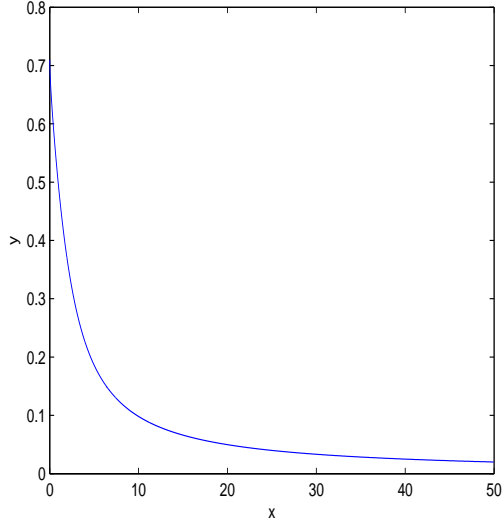}
\caption{\label{fig.1}The function  $y$ vs. $x$.}
\end{figure}
One can see numerical real and positive solutions to Eq. (22) in Table 1.
\begin{table}[ht]
\caption{Solutions to Eq. (22)}
\centering
\begin{tabular}{c c c c c c c c c c  c}\\[1ex]
\hline
$x$ & 1 & 2 & 3 & 4 & 5 & 6 & 7 & 8 & 9 & 10\\[0.5ex]
\hline
 $y$ & 0.491 & 0.364 & 0.280 & 0.224 & 0.186 & 0.158 & 0.137 & 0.121 & 0.108 & 0.098\\[0.5ex]
\hline
\end{tabular}
\end{table}
 As $r\rightarrow \infty$ the function $y(x)$ behaves as
\begin{equation}
y=\frac{1}{x}-\frac{2}{x^3}+{\cal O}(x^{-5}).
\label{23}
\end{equation}
Making use of Eqs. (21) and (23) one obtains the asymptotic of the electric field  as $r\rightarrow\infty$
\begin{equation}
E(r)=\frac{q_e}{r^2}-\frac{\beta q_e^3}{r^6}+{\cal O}(r^{-10}).
\label{24}
\end{equation}
Equation (24) shows that the correction to Coulomb's law is in the order of ${\cal O}(r^{-6})$.
At $\beta=0$ we have the Coulomb law $E=q_e/r^2$ of Maxwell's electrodynamics.

\section{The energy-momentum tensor and dilatation current }

By varying the action on the metric tensor $g_{\mu\nu}$ we obtain the symmetrical energy-momentum tensor
\begin{equation}
T_{\mu\nu}=\frac{2}{\sqrt{-g}}\frac{\partial (\sqrt{-g} {\cal L})}{\partial g^{\mu\nu}}.
\label{25}
\end{equation}
Making use of Eqs. (1) and (25) one finds the symmetrical energy-momentum tensor
\begin{equation}
T_{\mu\nu}=-\frac{(\beta{\cal F}+1)F_\mu^{~\alpha}F_{\nu\alpha}}{(1+2\beta{\cal F})^{3/2}}
+\frac{1}{2}\gamma{\cal G}(F_\mu^{~\alpha}\tilde{F}_{\nu\alpha}+F_\nu^{~\alpha}\tilde{F}_{\mu\alpha})
-g_{\mu\nu}{\cal L}.
\label{26}
\end{equation}
The trace of the energy-momentum tensor (26) is
\begin{equation}
{\cal T}\equiv T_\mu^{~\mu}=\frac{4\beta{\cal F}^2}{(1+2\beta{\cal F})^{3/2}}+2\gamma{\cal G}^2.
\label{27}
\end{equation}
In classical electrodynamics at $\beta=\gamma=0$ one arrives at the traceless energy-momentum tensor.
The dilatation current and its divergence are given by
\begin{equation}
D_{\mu}=x^\alpha T_{\mu\alpha},~~~~\partial_\mu D^{\mu}={\cal T}.
\label{28}
\end{equation}
Thus, the scale (dilatation) symmetry is broken as ${\cal T}\neq 0$ but in classical electrodynamics the dilatation symmetry holds.
The energy density, obtained from Eq. (26), is given by
\begin{equation}
\rho=T^{~0}_0=\frac{(1+\beta{\cal F})E^2}{(1+2\beta{\cal F})^{3/2}} +\frac{{\cal F}}{\sqrt{1+2\beta{\cal F}}}
+\frac{\gamma}{2}{\cal G}^2.
\label{29}
\end{equation}
The electric energy density at $\textbf{B}=0$, according to Eq. (29), is
\begin{equation}
\rho_E=T^{~0}_0=\frac{E^2}{2(1-\beta E^2)^{3/2}}.
\label{30}
\end{equation}
It worth noting that the energy density (30), according to Eq. (20), is infinite at $r=0$.
At $\textbf{E}=0$ the magnetic energy density becomes
\begin{equation}
\rho_M=T^{~0}_0=\frac{B^2}{2\sqrt{(1+\beta B^2)}}.
\label{31}
\end{equation}

\section{The dyonic solution}

To simplify the problem, in the following, we consider the case $\gamma=0$. When NED is the source of the gravity the action is
\begin{equation}
I=\int d^4x\sqrt{-g}\left(\frac{1}{16\pi G}R+ {\cal L}\right),
\label{32}
\end{equation}
where ${\cal L}$ is given by Eq. (1), $G$ is the Newton constant, $8\pi G\equiv M_{Pl}^{-2}$, and $M_{Pl}$ is the reduced Planck mass. The Einstein equation is given by
\begin{equation}
R_{\mu\nu}-\frac{1}{2}g_{\mu\nu}R=-8\pi GT_{\mu\nu}.
\label{33}
\end{equation}
The equation of motion for electromagnetic fields, obtained  by varying action (32) on electromagnetic potentials, is
\begin{equation}
\partial_\mu\left(\sqrt{-g}F^{\mu\nu}{\cal L}_{\cal F}\right)=0.
\label{34}
\end{equation}
We consider the metric to be the static and spherically symmetric, which is realized when $T_0^{~0}=T_r^{~r}$,
\begin{equation}
ds^2=-A(r)dt^2+\frac{1}{A(r)}dr^2+r^2(d\vartheta^2+\sin^2\vartheta d\phi^2),
\label{35}
\end{equation}
and the metric function is given by
\begin{equation}
A(r)=1-\frac{2M(r)G}{r}.
\label{36}
\end{equation}
The mass function for any asymptotically flat metrics is
\begin{equation}
M(r)=m_S+\int^r_0\rho(r)r^2dr=m_S+m_{el}-\int^\infty_r\rho(r)r^2dr,
\label{37}
\end{equation}
where $m_S$ is the Schwarzschild mass (the constant of integration in the Einstein equation) and $m_{el}=\int^\infty_0\rho(r)r^2dr$ is the electromagnetic mass.
In the following we consider BH solutions, where $m=m_S+m_{el}$ is the total mass of the BH.
The solutions of field equations are given by \cite{Bronnikov3}, \cite{Bronnikov4}
\begin{equation}
B^2=\frac{q^2_m}{r^4},~~~~E^2=\frac{q_e^2}{{\cal L}^2_{\cal F}r^4},
\label{38}
\end{equation}
and $q_m$ and $q_e$ are the magnetic and electric charges, respectively.
Making use of Eqs. (8) and (38) we obtain
\[
E^2=\frac{q_e^2(1+2\beta{\cal F})^3}{r^4(1+\beta{\cal F})^2},
\]
\begin{equation}
\beta{\cal F}=a-b\frac{(1+2\beta{\cal F})^3}{(1+\beta{\cal F})^2},~~~a=\frac{\beta q^2_m}{2r^4},~~~
b=\frac{\beta q_e^2}{2r^4},
\label{39}
\end{equation}
where $a$ and $b$ are unitless variables. It should be noted that for pure electric case ($q_m=0$, $\textbf{B}=0$) the solution to Eq. (39) coincides with the solution to Eq. (19) in flat space-time. Thus, there is the self-consistent solution in GR. From Eq. (39) one finds the cubic equation for $y\equiv\beta{\cal F}$
\begin{equation}
(8b+1)y^3+(12b+2-a)y^2+(6b+1-2a)y+b-a=0.
\label{40}
\end{equation}
The real solution to Eq. (40) corresponds to arbitrary magnetic and electric charges. Let us consider the simple self-dual solution with $q_e=q_m$ ($a=b$). Then from Eq. (40) we obtain the real solution $y=0$ (${\cal F}=0$, $E=B$) and two complex nonphysical solutions. From Eq. (29) one finds $\rho=E^2=B^2$, and making use of Eq. (37) and $E^2=q^2/r^4$ ($q\equiv q_e=q_m$), we obtain
\begin{equation}
M(r)=m-\frac{q^2}{r}.
\label{41}
\end{equation}
With the help of Eq. (36) one finds the metric function
\begin{equation}
A(r)=1-\frac{2mG}{r}+\frac{2q^2G}{r^2}.
\label{42}
\end{equation}
The metric function (42) represents the Reissner$-$Nordst\"{o}m (RN) solution, where $q^2_e+q^2_m=2q^2$. The similar result, in the self-dual case, holds in BI electrodynamics [39] and in logarithmic electrodynamics \cite{log}.

\section{The magnetic black hole}

Here, we consider the static magnetic BH. In this case ($q_e=0$) the invariant is ${\cal F}=q_m^2/(2r^4)$.
From Eq. (31) one finds the magnetic energy density
\begin{equation}
\rho_M=\frac{q_m^2}{2r^2\sqrt{r^4+\beta q_m^2}}.
\label{43}
\end{equation}
Making use of Eqs. (37) and (43) we obtain the mass function
\begin{equation}
M(r)=m-\frac{q_m^2}{2}\int_r^\infty \frac{dr}{\sqrt{r^4+\beta q_m^2}}.
\label{44}
\end{equation}
The integral in Eq. (44) represents the elliptic integral of the first kind.
We find the BH magnetic mass
\begin{equation}
m_M=\int_0^\infty\rho_M(r)r^2dr=\frac{2\Gamma^2(5/4)q_m^{3/2}}{\sqrt{\pi}\beta^{1/4}}\approx 0.927\frac{q_m^{3/2}}{\beta^{1/4}},
\label{45}
\end{equation}
where $\Gamma$ is the gamma-function. At $q_m=0$ we have $m_M=0$, and one comes to the Schwarzschild BH.
For large $r$ we can calculate the mass function (44) making use of the relation (as $r\rightarrow \infty$)
\begin{equation}
\frac{1}{\sqrt{r^4+\beta q_m^2}}=\frac{1}{r^2}-\frac{\beta q_m^2}{2r^6}+\frac{3\beta^2 q_m^4}{8r^{10}}+{\cal O}(r^{-13}).
\label{46}
\end{equation}
Taking into account Eqs. (44) and (46) we obtain the mass function as $r\rightarrow \infty$
\begin{equation}
M(r)=m-\frac{q_m^2}{2r}+\frac{\beta q_m^4}{20r^5}-\frac{\beta^2 q_m^6}{48r^9}+{\cal O}(r^{-12}).
\label{47}
\end{equation}
Making use of Eqs. (36) and (47) we find the metric function as $r\rightarrow \infty$
\begin{equation}
A(r)=1-\frac{2mG}{r}+\frac{q_m^2G}{r^2}-\frac{\beta q_m^4G}{10r^6}+\frac{\beta^2 q_m^6G}{24r^{10}}+{\cal O}(r^{-13}).
\label{48}
\end{equation}
In accordance with Eq. (48), the correction to the RN solution is in the order of ${\cal O}(r^{-6})$.
Introducing the unitless variable $y=r/(\sqrt{q_m}\beta^{1/4})$, and using Eq. (45), we represent Eq. (44) as follows:
\[
M(y)=m-\frac{q_m^{3/2}}{2\beta^{1/4}}\int_y^\infty \frac{dy}{\sqrt{y^4+1}}
\]
\begin{equation}
=m_S-\frac{q_m^{3/2}}{2\beta^{1/4}}\left(\sqrt[4]{-1}F(i\sinh^{-1}(\sqrt[4]{-1}y)|-1)\right),
\label{49}
\end{equation}
where $F(\varphi|k^2)$ is the elliptic function of the first kind and $\sinh^{-1}(x)$ is the inverse hyperbolic sinh-function. By introducing new unitless constants $C=m_S\beta^{1/4}/q_m^{3/2}$, $B=q_mG/\sqrt{\beta}$, we obtain the metric function (36) for arbitrary $r$ as follows:
\begin{equation}
A(y)=1-\frac{B}{y}\left(2C-\sqrt[4]{-1}F(i\sinh^{-1}(\sqrt[4]{-1}y)|-1)\right).
\label{50}
\end{equation}
The event horizon radii ($y_+=r_+/(\sqrt{q_m}\beta^{1/4})$) for different parameters $C$ ($B=1$) are represented in Table 2.
\begin{table}[ht]
\caption{Horizon radii ($B=1$)}
\centering
\begin{tabular}{c c c c c c c c c c  c}\\[1ex]
\hline
$C$ & 0.05 & 0.1 & 0.15 & 0.2 & 0.25 & 0.3 & 0.35 & 0.4 & 0.45 & 0.5 \\[0.5ex]
\hline
 $y_+$ & 1.08 & 1.32 & 1.50 & 1.66 & 1.81 & 1.94 & 2.07 & 2.20 & 2.33 & 2.45 \\[0.5ex]
\hline
\end{tabular}
\end{table}
The metric function (50) is depicted in Fig. 2.
\begin{figure}[h]
\includegraphics[height=3.0in,width=3.0in]{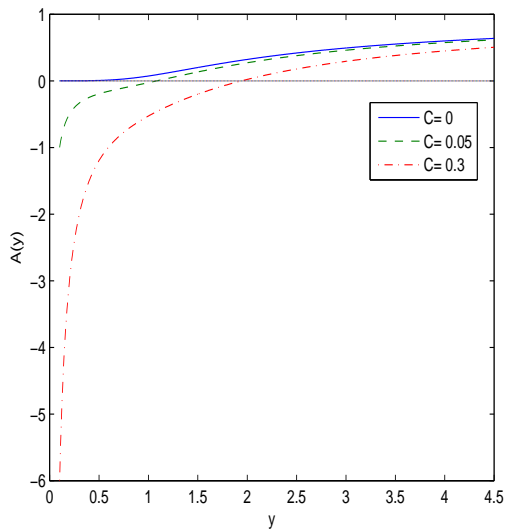}
\caption{\label{fig.2} The plot of the function $A(y)$ for $B=1$. The dashed curve corresponds to $C=0.05$, the solid curve is for $C=0$, and the dash-doted curve corresponds to $C=0.3$.}
\end{figure}
According to Fig. 2 there are (at $B=1$) BH solutions with one horizon (see Table 2).

We obtain the Ricci scalar from Eqs. (27) and (33) at $E=0$,
\begin{equation}
R=8\pi G{\cal T}=\frac{8\pi G\beta q_m^4}{r^2(r^4+\beta q_m^2)^{3/2}}.
\label{51}
\end{equation}
As $r\rightarrow \infty$ the Ricci scalar goes to zero, $R\rightarrow 0$, and space-time becomes flat. But as $r\rightarrow 0$ the curvature possesses a singularity.

\section{The black hole thermodynamics}

Let us study the black holes thermodynamics and the thermal stability of magnetic BHs. The Hawking temperature is defined as
\begin{equation}
T_H=\frac{\kappa}{2\pi}=\frac{A'(r_+)}{4\pi},
\label{52}
\end{equation}
where $\kappa$ is the surface gravity and $r_+$ is the event horizon radius. Making use of Eqs. (36) and (37) we obtain the relations
\begin{equation}
A'(r)=\frac{2 GM(r)}{r^2}-\frac{2GM'(r)}{r},~~~M'(r)=r^2\rho,~~~M(r_+)=\frac{r_+}{2G}.
\label{53}
\end{equation}
From Eqs. (52), and (53) one finds the Hawking temperature
\begin{equation}
T_H=\frac{1}{4\pi\sqrt{q_m}\beta^{1/4}}\biggl(\frac{1}{y_+}-\frac{q_mG}{\sqrt{\beta}y_+\sqrt{y_+^4+1}}\biggr).
\label{54}
\end{equation}
Making use of Eq. (50) and the equation $A(y_+)=0$ we obtain
\[
B\equiv\frac{q_mG}{\sqrt{\beta}}=\frac{y_+}{D},
\]
\begin{equation}
D\equiv (2C-\sqrt[4]{-1}F(i\sinh^{-1}(\sqrt[4]{-1}y_+)|-1)).
\label{55}
\end{equation}
Substituting Eq. (55) into Eq. (54) we arrive at the equation for the Hawking temperature
\begin{equation}
T_H=\frac{1}{4\pi\sqrt{q_m}\beta^{1/4}}\biggl(\frac{1}{y_+}-\frac{1}{D\sqrt{y_+^4+1}}\biggr).
\label{56}
\end{equation}
The plots of the functions $T_H(y_+)\sqrt{q_m}\beta^{1/4}$ for different parameters $C$ are given in Fig. 3.
\begin{figure}[h]
\includegraphics[height=3.0in,width=3.0in]{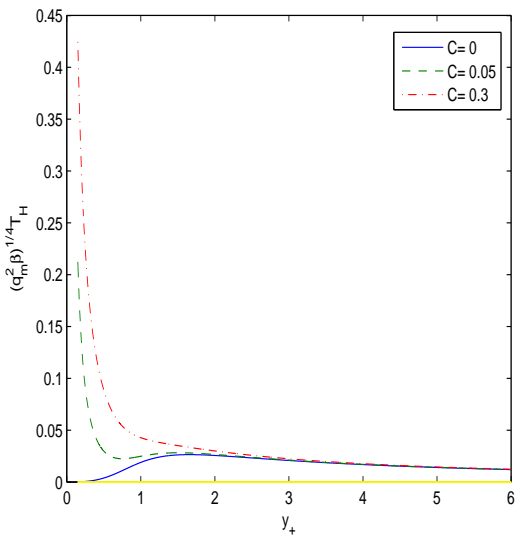}
\caption{\label{fig.3}The plot of the function $T_H\sqrt{q_m}\beta^{1/4}$ vs $y_+$. The dashed curve corresponds to $C=0.05$, the solid curve is for $C=0$, and the dash-doted curve corresponds to $C=0.3$.}
\end{figure}
In accordance with Fig. 3 the Hawking temperature is positive for any values of $C\geq 0$. The case $C=0$ corresponds to the BH with zero Schwarzschild mass ($m_S=0$). Figure 3 shows that there are maximums in the Hawking temperature (the Davies points) at small values of the Schwarzschild mass ($C=0$ and $C=0.05$), and as a result, phase transitions occur.

The different stability phases of the BH can be studied defining the signs of the heat
capacity and the Helmholtz free energy \cite{Page}. From the  Hawking entropy of the BH $S=\mbox{Area}/(4G)=\pi r_+^2/G=\pi y_+^2q_m\sqrt{\beta}/G$ we obtain the heat capacity
\begin{equation}
C_q=T_H\left(\frac{\partial S}{\partial T_H}\right)_q=\frac{T_H\partial S/\partial y_+}{\partial T_H/\partial y_+}=\frac{2\pi q_m\sqrt{\beta}y_+T_H}{G\partial T_H/\partial y_+}.
\label{57}
\end{equation}
In accordance with Eq. (57) the heat capacity diverges if the Hawking temperature has the extremum, $\partial T_H/\partial y_+=0$. Making use of Eqs. (56) and (57) we find the heat capacity in the terms of unitless variables
\begin{equation}
\frac{G}{q_m\sqrt{\beta}}C_q=\frac{2\pi y_+(y_+^4+1)D\left(D\sqrt{y_+^4+1}-y_+\right)}{y_+^2\sqrt{y_+^4+1}-(y_+^4+1)^{3/2}D^2+2y_+^5D}.
\label{58}
\end{equation}
\begin{figure}[h]
\includegraphics[height=3.0in,width=3.0in]{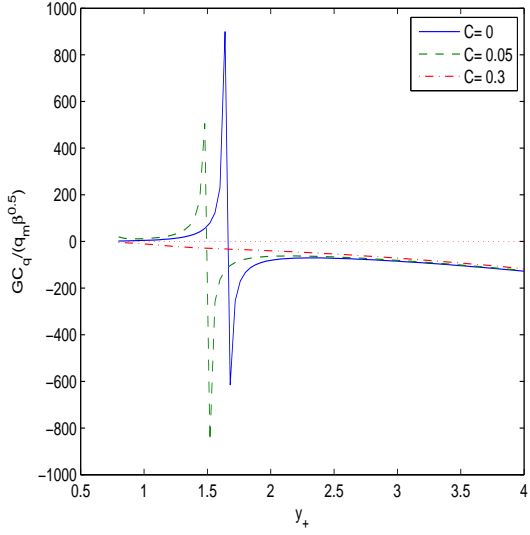}
\caption{\label{fig.4}The plot of the function $GC_q/(q_m\sqrt{\beta})$ vs $y_+$. The dashed curve corresponds to $C=0.05$, the solid curve is for $C=0$, and the dash-doted curve corresponds to $C=0.3$.}
\end{figure}
In Fig. 4 the heat capacities versus the variable $y_+$ for different parameters $C$ are depicted. Singularities  in the heat capacity for $C=0$ and $C=0.05$ show the points ($y_+\approx 1.65$ and $y_+\approx1.5$) where the second-order phase transitions occur. These horizon radii ($r_+=y_+\sqrt[4]{\beta}\sqrt{q_m}$) correspond to the maximums of the Hawking temperatures (see Fig. 3).
The heat capacity in these points is changed from negative infinity to positive infinity. The whole evaporation process in the discontinuity points is separated by the early stage with negative heat capacity and the late stage with positive heat capacity at the maximum temperature. The temperature increases, for the early evaporation process, because the mass of the BH decreases. This process starts from initial unstable large BH and ends at the final stable BH. When the heat capacity is negative the BHs are unstable because of the BH evaporation. The heat capacity for $C=0.3$ is negative because the slope in the temperature graph is negative (see Fig. 3). This shows the Schwarzschild behaviour of the heat capacity.

To compute the Helmholtz free energy we use the expression as follows:
\begin{equation}
F=m-T_HS,
\label{59}
\end{equation}
where the mass of the BH $m$ plays the role of the internal energy of the BH, the Hawking temperature is $T_H$ and the entropy reads $S=\pi r_+^2/G$. The Bekenstein-Hawking entropy used should coincide with other definitions of the entropy because the BH entropy is a Noether charge \cite{Wald}. From Eqs. (56), (59), and using the definitions $y=r/(\sqrt{q_m}\sqrt[4]{\beta})$, $C=m_S\sqrt[4]{\beta}/q_m^{3/2}$, $B=q_mG/\sqrt{\beta}$, we obtain
\begin{equation}
\frac{GF}{\sqrt{q_m}\beta^{1/4}}=BC-\frac{1}{4}\biggl(y_+-\frac{y_+^2}{D\sqrt{y_+^4+1}}\biggr).
\label{60}
\end{equation}
Replacing Eq. (55) into Eq. (60) one finds
\begin{equation}
\frac{GF}{\sqrt{q_m}\beta^{1/4}}=\frac{y_+\biggl(4C\sqrt{y_+^4+1}+y_+\biggr)}{4D\sqrt{y_+^4+1}}-\frac{y_+}{4}.
\label{61}
\end{equation}
Here, we have introduced the unitless reduced free energy $GF/(\sqrt{q_m}\beta^{1/4})$. The plots of the function (61) for different values of $C$ are given in Fig. 5.
\begin{figure}[h]
\includegraphics[height=3.0in,width=3.0in]{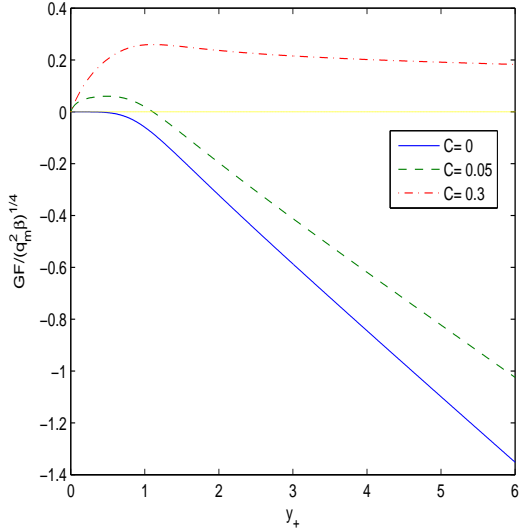}
\caption{\label{fig.5}The plot of the function $GF/(\sqrt{q_m}\beta^{1/4})$ vs $y_+$. The dashed curve corresponds to $C=0.1$, the solid curve is for $C=0$, and the dash-doted curve corresponds to $C=0.3$.}
\end{figure}
The BHs with $F > 0$ will be tunneling to decay for the pure radiation, but the BHs with $F < 0$ are stable.
Figure 5, as well as Fig. 4, show that the BHs with $C=m_S\sqrt[4]{\beta}/q_m^{3/2}= 0.3$ (and for $C>0.3$) are unstable because the Helmholtz free energy $F >0$ and the heat capacity $C_q < 0$. At the corresponding Schwarzschild mass $m_S$, BHs are unstable because of the BH evaporation. The phase at $F <0$, $C_q > 0$, with the stable BHs, is realized for zero Schwarzschild mass ($C=0$) at $y_+<1.65$.
In accordance with Figs. 4 and 5, there are other phases with $F > 0$, $C_q > 0$ (the unstable phase with $C=0.05$, $y_+<1.5$) and $F <0$, $C_q < 0$ ($C=0$, $y_+>1.65$ and for $C=0.05$, $y_+>1.5$). In the case of $F < 0$, $C_q < 0$  the BHs are less energetic than the pure radiation and therefore BHs do not decay through tunneling. Since the heat capacities are negative, the BH will increase its temperature with decreasing the mass of BHs. Such phases also hold in another model \cite{Jarillo}.

 \section{Conclusion}

A new model of NED with two parameters $\beta$ and $\gamma$ is proposed. For weak fields our model is converted to Maxwell's electrodynamics so that the correspondence principle takes place. When $\gamma\neq 2\beta$ the birefringence effect occurs but in the case $\gamma = 2\beta$ there is no the phenomenon of the birefringence like to classical and BI electrodynamics. It is worth noting that in QED, due to quantum loop corrections, the birefringence effect holds. It was shown that as $1\geq\beta{\cal F}\geq -1$ the principles of causality and unitarity occur. It was demonstrated that the dual symmetry is broken in our model as well as in QED. We shown that the singularity of the electric field at the origin of point-like particles is absent and the maximum electric field in the center is $E(0)=1/\sqrt{\beta}$. In the case of the electrostatics, the correction to the Coulomb law as $r\rightarrow \infty$ is in the order of ${\cal O}(r^{-6})$. It was shown that the scale (dilatation) symmetry is broken because of the dimensional parameters $\beta$ and $\gamma$.

We studied the dyonic and magnetic BHs in GR and found solutions and asymptotics as $r\rightarrow \infty$. In the self-dual case ($q_e=q_m$) the corrections to Coulomb's law and RN solutions are absent. The magnetic mass of the BH was calculated which is finite. The mass and metric functions were calculated.

The thermal stability of BHs was studied. We calculated the Hawking temperature, the heat capacity and the Helmholtz free energy of BHs. It was shown that at some parameters $C$ (or $m_S$) and event horizon radii BHs are stable or unstable. We demonstrated that the heat capacity diverges at some event horizon radii $r_+$ ($y_+$) and the phase transitions of the second-order take place.
We discovered a new stability region of BH solutions when the heat capacity and the Helmholtz free energy are negative. In this case BHs are less energetic than the pure radiation and BHs do not decay via tunneling.
The model proposed is of theoretical interest because of its simplicity and the presence of different thermodynamics phases of BHs.


\begin{thebibliography}{99}

\bibitem{Shapere} A. D. Shapere, S. Trivedi, and F. Wilczek, Mod. Phys. Lett. A \textbf{6}, 2677 (1991).
\bibitem{Mignemi0}S. Mignemi, Phys. Rev. D \textbf{51}, 934 (1995).
\bibitem{Cvetic} M. Cvetic and A. A. Tseytlin, Phys. Rev. D \textbf{53}, 5619 (1996); Erratum: Phys. Rev. D \textbf{55}, 3907 (1997).
\bibitem{Jatkar}D. P. Jatkar, S. Mukherji, and S. Panda, Nucl. Phys. B \textbf{484}, 223 (1997).
\bibitem{Chamseddine}A. H. Chamseddine and W. A. Sabra, Phys. Lett. B \textbf{485}, 301 (2000).
\bibitem{Chow}D. D. K. Chow and G. Compere, Phys. Rev. D \textbf{89}, 065003 (2014).
\bibitem{Meessen}P. Meessen, T. Ortín, and P. F. Ramírez, JHEP \textbf{1710}, 066 (2017).
\bibitem{Pang}H. Lu, Y. Pang, and C.N. Pope, JHEP \textbf{1311}, 033 (2013).
\bibitem{Hartnoll}S. A. Hartnoll and P. Kovtun, Phys. Rev. D \textbf{76}, 066001 (2007).
\bibitem{Hartnoll1}S. A. Hartnoll, P. K. Kovtun , M. Muller, and S. Sachdev, Phys. Rev. B \textbf{76}, 144502 (2007).
\bibitem{Dutta}S. Dutta, A. Jain, and R. Soni, JHEP \textbf{1312}, 060 (2013).
\bibitem{Born} M. Born and L. Infeld, Proc. R. Soc. Lond. \textbf{144}, 425 (1934).
\bibitem{Heisenberg} W. Heisenberg and H. Euler, Z. Phys. \textbf{98}, 714 (1936).
\bibitem{Soleng} H. H. Soleng, Phys. Rev. D \textbf{52}, 6178 (1995).
\bibitem{Shabad} D. M. Gitman and A. E. Shabad, Eur. Phys. J. C \textbf{74}, 3186 (2014).
\bibitem{Shabad1} C. V. Costa, D. M. Gitman, and A. E. Shabad, Phys. Scripta \textbf{90}, 074012 (2015).
 \bibitem{Krug1} S. I. Kruglov, Commun. Theor. Phys. \textbf{66}, 59 (2016).
 \bibitem{Krug2}S. I. Kruglov, Ann. Phys. \textbf{353}, 299 (2015).
\bibitem{Kruglov1}S. I. Kruglov, Mod. Phys. Lett. A \textbf{32}, 1750201 (2017).
\bibitem{Pellicer} R. Pellicer and R. J. Torrence, J. Math. Phys. \textbf{10}, 1718 (1969).
\bibitem{Oliveira} H. P. de Oliveira, Class. Quant. Grav. \textbf{11}, 1469 (1994).
\bibitem{Ayon1} E. Ay\'{o}n-Beato and A. Gar\'{c}ia, Phys. Rev. Lett.  \textbf{80}, 5056 (1998).
\bibitem{Bronnikov0} K. A. Bronnikov, V. N. Melnikov, G. N. Shikin, and K. P. Staniukovich, Ann. Phys.  \textbf{118}, 84 (1979).
\bibitem{Bronnikov} K. A. Bronnikov, Phys. Rev. D \textbf{63}, 044005 (2001).
\bibitem{Bronnikov1}  K. A. Bronnikov, Phys. Rev. Lett. \textbf{85}, 4641 (2000).
 \bibitem{Bronnikov2}   K. A. Bronnikov, G. N. Shikin, and E. N. Sibileva, Grav. Cosmol. \textbf{9}, 169 (2003).
\bibitem{Burinskii} A. Burinskii and S. R. Hildebrandt, Phys. Rev. D \textbf{65}, 104017 (2002).
\bibitem{Diaz} J. Diaz-Alonso and D. Rubiera-Garcia, Phys. Rev. D \textbf{81}, 064021 (2010).
\bibitem{Breton} N. Breton, Gen. Rel. Grav. \textbf{ 37}, 643 (2005).
\bibitem{Novello} M. Novello, S. E. Perez Bergliaffa, and J. M. Salim, Class. Quant. Grav. \textbf{17}, 3821 (2000).
\bibitem{Quiros} R. Garcia-Salcedo, T. Gonzalez, and I. Quiros, Phys. Rev. D \textbf{89}, 084047 (2014).
\bibitem{Hendi1} S. H. Hendi, Ann. Phys. \textbf{333}, 282 (2013).
\bibitem{Lemos} J. P. S. Lemos and V. T. Zanchin, Phys. Rev. D \textbf{83}, 124005 (2011).
\bibitem{Myung} Y. S. Myung, Y.-W. Kim, and Y.-J. Park, Gen. Rel. Grav. \textbf{41}, 1051 (2009).
\bibitem{Balart} L. Balart and E. C. Vagenas, Phys. Rev. D \textbf{90}, 124045 (2014).
\bibitem{Krug4} S. I. Kruglov, Phys. Rev. D \textbf{94}, 044026 (2016).
\bibitem{Krug6}S. I. Kruglov, Ann. Phys. (Berlin) \textbf{528}, 588 (2016).
\bibitem{Yajima} H. Yajima and T. Tamaki, Phys. Rev. D \textbf{63}, 064007 (2001).
\bibitem{Bronnikov3} K. A. Bronnikov, Grav. Cosmol. \textbf{23}, 343 (2017).
\bibitem{Bronnikov4} K. A. Bronnikov, Int. J. Mod. Phys. D \textbf{27}, 1841005 (2018).
\bibitem{Krug9}S. I. Kruglov, Int. J. Mod. Phys. A \textbf{33}, 1850023 (2018).
\bibitem{Kruglov3}S. I. Kruglov, Ann. Phys. \textbf{383}, 550 (2017).
\bibitem{log}S. I. Kruglov, Grav. Cosmol. \textbf{25}, 190 (2019).
\bibitem{Garcia} R. Garc\'{i}a-Salcedo and N. Breton, Int. J. Mod. Phys. A \textbf{15}, 4341 (2000).
\bibitem{Camara} C. S. Camara, M. R. de Garcia Maia, J. C. Carvalho, and J. A. S. Lima, Phys. Rev. D \textbf{69}, 123504 (2004).
 \bibitem{Elizalde} E. Elizalde, J. E. Lidsey, S. Nojiri, and S. D. Odintsov, Phys. Lett. B \textbf{574}, 1 (2003).
\bibitem{Novello1} M. Novello, S. E. Perez Bergliaffa, and J. M. Salim, Phys. Rev. D \textbf{69}, 127301 (2004).
\bibitem{Novello2} M. Novello, E. Goulart, J. M. Salim, and S. E. Perez Bergliaffa, Class. Quant. Grav. \textbf{24}, 3021 (2007).
\bibitem{Vollick} D. N. Vollick, Phys. Rev. D \textbf{78}, 063524 (2008).
\bibitem{Krug12}S. I. Kruglov, Phys. Rev. D \textbf{92},  123523 (2015).
\bibitem{Krug13}S. I. Kruglov, Int. J. Mod. Phys. A \textbf{32}, 1750071 (2017).
\bibitem{Kruglov4}S. I. Kruglov, Int. J. Mod. Phys. D \textbf{25}, 1640002 (2016).
\bibitem{Rizzo} A. Cadene, P. Berceau, M. Fouche, R. Battesti, and C. Rizzo, Eur. Phys. J. D \textbf{68}, 16 (2014).
\bibitem{Valle} F. Della Valle, et al, Phys. Rev. D \textbf{90} 092003 (2014).
\bibitem{Battesti} R. Battesti and C. Rizzo, Rep. Prog. Phys. \textbf{76}, 016401 (2013).
\bibitem{Krug} S. I. Kruglov, J. Phys. A \textbf{43}, 375402 (2010).
\bibitem{Kruglov9} S. I. Kruglov, Phys. Rev. D \textbf{75}, 117301 (2007).
\bibitem{Shabad2}A. E. Shabad and V. V. Usov, Phys. Rev. D \textbf{83}, 105006 (2011).
\bibitem{Hehl} F. W. Hehl and Yu. N. Obukhov, \textit{Foundations of classical electrodynamics: Chage, flux, and metric} (Birkh\"{a}user, Boston, 2003).
\bibitem{Gibbons} G. W. Gibbons and D. Rasheed,  Nucl. Phys. B \textbf{454} (1995) 185.
\bibitem{Page} S. W. Hawking and D. N. Page, Commun. Math. Phys. \textbf{87}, 577 (1983).
\bibitem{Wald}R. M. Wald, Phys. Rev. D \textbf{48}, R3427 (1993).
\bibitem{Jarillo}J. A. R. Cembranos, A. Cruz-Dombriz, and J. Jarillo, Universe, \textbf{1}, 412 (2015).

\end{thebibliography}
\end{document}